# Inelastic scattering puts in question recent claims of Anderson localization of light

**To the Editor —** In a recent Letter in *Nature Photonics*, Sperling et al.[1] reported the observation of Anderson localization of light in three dimensions. In contrast to previous studies, the authors designed their experiment to be insensitive to absorption. To this end, they measured the time-dependent broadening of a high-intensity, short laser pulse transmitted through a highly scattering medium made from compressed $TiO_2$ powder. In analogy with the case of disordered optical fibres[2], localized light is expected to be laterally confined to roughly the localization length $\xi$. Using an ultrafast imaging system, Sperling et al. experimentally observed the saturation of the time-dependent transverse width of the total transmitted light intensity, and from this, they derived the claim for the first unequivocal observation of the three-dimensional localization of light.

In this correspondence, we would like to point out that the Letter of Sperling et al. does not report on the observation of elastic scattering of light waves, which is considered to be a necessary condition for the occurrence of Anderson localization. In his recent PhD thesis[3], Wolfgang Bührer (who was supervised by Maret and Aegerter) reported that highly nonlinear contributions exceed the linear (elastic) scattering signal by at least one order of magnitude in the most relevant regime, namely the long-time regime ($\tau/\tau_{max} \approx 3$ in Fig. 2 of ref. 1). Although such incoherent light makes an extremely small contribution to the total transmission, it becomes the dominant contribution to the late arriving signal once the elastically scattered light has leaked out of the sample. Wavelength-resolved experiments reported in this thesis on samples similar or identical to the ones studied in ref. 1, show that the non-exponential tail of the transmitted pulse disappears when the spectrally shifted contributions are blocked by a band-pass filter.

We believe that the dominant contribution of incoherent light puts in question not only the recent claims by Sperling et al. but also similar claims of localization by the same group in 2006 based on time-resolved measurements alone[4]. The long-time regime lies at the heart of both claims of localization, as this is the regime in which the saturation of the transverse width was observed in ref. 1 and in which the deviation from non-exponential decay was observed in ref. 4. Although Bührer extensively studied this incoherent contribution previously[3], it was not mentioned in ref. 1.

We note that it is relatively easy to confuse nonlinear effects with localization in this type of experiment using pulsed laser sources. For example, photons generated by nonlinear processes (such as radiative decay after two-photon absorption) are emitted with a distribution of time delays $\Delta t$, which contribute to a narrower transmission profile

$T(\rho, t) = T_{elastic}(\rho, t) + T_{inelastic}(\rho, t - \Delta t)$ at the output. The transition from elastic to inelastic scattering in the long-time regime can thus result in an 'apparent' saturation of the transverse width, resembling that of localization. Also, the different particle sizes and the different pressures used to produce samples of varying packing fractions can lead to differences in nonlinear optical coefficients, which can be easily misinterpreted as a localization effect that depends on the scattering strength ($kl^*$), whereas it is actually a nonlinear optical effect. ❑

Frank Scheffold[1] and Diederik Wiersma[2,3]
[1]Physics Department, University of Fribourg, Chemin du Musée 3, 1700 Fribourg, Switzerland, [2]European Laboratory for Non-linear Spectroscopy (LENS), University of Florence, Via N. Carrara 1, 50019 Sesto Fiorentino, Firenze, Italy, [3]Istituto Nazionale di Ottica (CNR-INO), Largo Fermi 6, 50125 Firenze, Italy.
e-mail: frank.scheffold@unifr.ch, wiersma@lens.unifi.it

---

**Maret et al. reply:** The interplay between nonlinear effects and Anderson localization in disordered optical fibres[1] has recently attracted great interest, and it is important in the action of random lasers in which closed multiple scattering loops have enhanced intensity[2]. As optical nonlinearities in $TiO_2$ can give valuable information on the nature of light transport in strongly scattering powders, we studied these effects in an extended experimental and theoretical investigation (to be published). Now, Scheffold and Wiersma have put forward that such effects may question the interpretation of the results of our recent experiments in terms of Anderson localization[3,4].

As noted by Scheffold and Wiersma, our measurements of the time-dependent transmission profiles[4] eliminate the influence of absorption. This technique thus reveals the signatures of Anderson localization more clearly than integrated transmission data[3]. The relevant signals appear at different timescales, which is also true for the nonlinear contributions in our study to be published. Based on the arguments put forth by Scheffold and Wiersma, one would expect that no saturation of the profile would be observed when a band-pass filter for the incoming wavelength is inserted. However, such an experiment (Fig. 1) clearly shows saturation of the profile width, similar to that reported in ref. 4, irrespective of the detected frequency.

In addition, we emphasize that the dependence of the localization length on $kl^*$ was determined in two separate ways in ref. 4. Besides the variation of the samples and the packing fractions mentioned by Scheffold and Wiersma, we also investigated the change in the scattering strength induced by varying the incident wavelength. Our data presented in two different studies[3,4], exhibit a remarkable scaling with the turbidity measured by $kl^*$ — exactlZ as expected for three-dimensional Anderson localization.

Even if, in an extraordinary coincidence, the change in nonlinear effects mentioned by Scheffold and Wiersma scaled exactly as *kl\**, an alternative method of changing *kl\** would not give the same quantitative results, whereas we have observed that it does[4]. Also, the determination of the localization length in two experiments that are conceptually very different[3,4] gives the same results (within errors) as a function of *kl\**.

Although in different experiments[3,4], the input power density varied by more than four orders of magnitude, the features in the time-dependent integrated transmission varied by less than a factor of five at long times, showing that the nonlinear contributions are weak. Moreover, the nonlinear scenario invoked by Scheffold and Wiersma to explain the narrowing of the observed profiles would not cause saturation at long times, rather they would only reduce the rate of increase, which has not been observed. In that case, the relative contributions should be strongly dependent on the incident intensity, which again has not been observed.

In conclusion, elastically scattered light does show a plateauing width in our experiments, and the observation of a plateau clearly scales with *kl\**, as determined by two independent means. Small nonlinear contributions via phonon-assisted processes do exist and provide density of states inside the electronic bandgap of $TiO_2$, which can qualitatively account for the observed frequency shifts given the intensity enhancement in the localized modes. Both integrated time-dependent transmission and transmission profiles are quantitatively accounted for by the self-consistent Vollhardt–Wölfle theory of Anderson localization of photons in the presence of realistic nonlinear effects, boundaries and other sample parameters[5] matched to the experiment (see Fig. 1). Thus, although small nonlinear effects may eventually be present, they do not invalidate the experimental observation of the Anderson localization transition with light. ❑

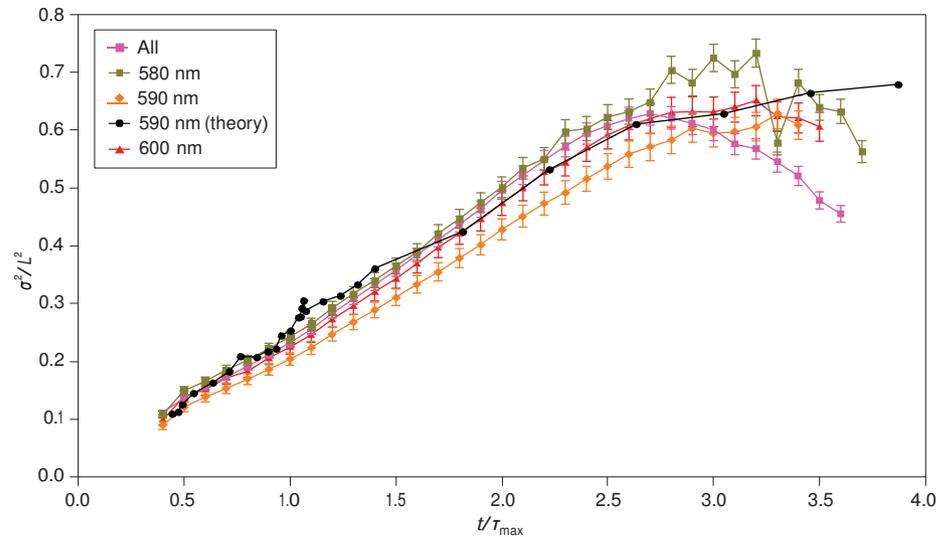

**Figure 1 | Time dependence of the mean square width** $\sigma^2$ normalized to sample thickness *L* for pressed powder of a strongly scattering sample (R700, see ref. 4, *kl\** = 2.7(3), *L* = 0.86 mm, absorption time = 0.9(1) ns) at incident laser wavelength 590 nm and different band-pass filters between the sample and the detector. The legend shows the centre filter wavelengths; the filter width was 10 nm. Coloured lines represent experimentally measured data, whereas the black line indicates curve predicted by theory for the same sample parameters as those used in the experiment.

Georg Maret[1], Tilo Sperling[1], Wolfgang Bührer[1], Andreas Lubatsch[2], Regine Frank[3] and Christof M. Aegerter[4]

[1]Fachbereich Physik, Universität Konstanz, Universitätsstrasse 10, 78457 Konstanz, Germany, [2]Electrical Engineering, Precision Engineering, Information Technology, Georg-Simon-Ohm University of Applied Sciences, Kesslerplatz 12, 90489 Nürnberg, Germany, [3]Institut for Theoretical Physics, Auf der Morgenstelle 14, Eberhard-Karls-University, 72076 Tübingen, Germany, [4]Physik-Institut, University of Zürich, 8057 Zürich, Switzerland.
*e-mail: aegerter@physik.uzh.ch